\documentclass[twocolumn]{aa}  
\usepackage{graphicx}
\usepackage{txfonts}
\usepackage{natbib}
\bibpunct{(}{)}{;}{a}{}{,}
\begin{document}
   \title{Reaching micro-arcsecond astrometry with long baseline optical interferometry}

   \subtitle{Application to the GRAVITY instrument}

\titlerunning{Reaching micro-arcsecond astrometry with GRAVITY}

   \author{S. Lacour\inst{1}\and
          F. Eisenhauer\inst{2}\and
          S. Gillessen\inst{2}\and
          O. Pfuhl\inst{2} \and 
          J. Woillez\inst{3} \and 
          H. Bonnet\inst{3} \and 
          G. Perrin\inst{1}\and
          B. Lazareff\inst{4} \and
          S. Rabien\inst{2}\and
          V. Lapeyr\`ere\inst{1}\and
          Y. Cl\'enet\inst{1}\and 
          P. Kervella\inst{1}\and 
          Y. Kok\inst{2}
                    }

   \institute{
   LESIA/Observatoire de Paris, CNRS, UPMC, Universit\'e Paris Diderot, 5 place Jules Janssen, 92195 Meudon,
France 
	  \and
	  Max-Planck Institut fur extraterrestrische Physik (MPE), Giessenbachstrasse 1, 85748, Garching, Germany
	  \and
  European Southern Observatory (ESO), Karl-Schwarzschild-Str. 2, 85748, Garching, Germany
  \and
  Institut de Plan\'etologie et d'Astrophysique de Grenoble, 38400 St Martin d'H\`eres, France
             }

   \date{Received September 15, 1996; accepted March 16, 1997}

 
  \abstract
   {A basic principle of long baseline interferometry is that an optical path difference (OPD) directly translates into an astrometric measurement. In the simplest case, the OPD is equal to the scalar product between the vector linking the two telescopes and the normalized vector pointing toward the star. However, in some circumstances, a too simple interpretation of this scalar product leads to seemingly conflicting results, called here "the baseline paradox".}
   { For micro-arcsecond accuracy astrometry, we have to model in full the  metrology measurement. It involves a complex system subject to many optical effects: from pure baseline errors to static, quasi-static and high order optical aberrations. The goal of this paper is to present the strategy used by the "General Relativity Analysis via VLT InTerferometrY" instrument (GRAVITY) to minimize the biases introduced by these defects.}
   {It is possible to give an analytical formula on how the baselines and tip-tilt errors  affect the astrometric measurement. This formula depends on the limit-points of three type of baselines: the wide-angle baseline, the narrow-angle baseline, and the imaging baseline. We also, numerically, include non-common path higher-order aberrations, whose amplitude were measured during technical time at the Very Large Telescope Interferometer (VLTI). We end by simulating the influence of high-order common-path aberrations due to atmospheric residuals calculated from a Monte-Carlo simulation tool for Adaptive optics (AO) systems.}
   {The result of this work is an error budget of the biases caused by the multiple optical imperfections, including optical dispersion. We show that the beam stabilization through both focal and pupil tracking is crucial to the GRAVITY system. Assuming the instrument pupil is stabilized at a 4\ cm level on M1, and a field tracking below $0.2\,\lambda/D$, we show that GRAVITY will be able to reach its objective of $10\ \mu$as accuracy.}
   {}

   \keywords{astrometry, instrumentation: interferometers, instrumentation: high angular resolution}

   \maketitle
%

\section{Introduction}

A particular interest of astronomical long baseline optical interferometers is their ability to perform high accuracy angular astrometry \citep{1992A&A...262..353S}. The basic idea is that they can leverage on the high resolution offered by baselines of several hundred of meters. However, the implementation of astrometry is not straightforward. The practical measurement is a monitoring of the paths of the stellar lights inside the facility. More precisely, it is a measurement of the difference in optical path (OP) length between each arm of the interferometer. This is usually done with an internal metrology system, going from the interferometric lab  up to the telescopes.
The idea behind this is that an homodyne interferometer converts an angular position on the sky to
an optical path difference (OPD). In the first order approximation, this optical path difference is equal to the direction of the star in the sky
($\vec
s$) projected onto the baseline vector of the interferometer ($\vec B$):
\begin{equation}
 OPD=\vec s \cdot \vec B\,.
 \label{eq1}
\end{equation}
Within this simple equation is hidden a lot of complexity. Above all, how is the baseline vector defined? The intuitive answer
is to define the baseline by the vector which links the geographical position of the two telescopes. But unfortunately,
it only has a sense in the case of two perfectly identical telescopes. This immediately implies some questions :
what happens when the two telescopes are not identical? What happens if, for one reason or another, the optical alignment of the two telescopes is not identical?

When dealing with precision astrometry, these kind of problem  gets pivotal.  
The main scientific objective of the GRAVITY instrument is to probe the event horizon of the Galactic Center black hole \citep{2011Msngr.143...16E}. It will have to reach an unprecedented astrometric accuracy between two
one-arcsec-separated objects of the order of $10\,\mu$as (ten part par million). According to Eq.~(\ref{eq1}),
it means that the baseline of the Very Large Telescope Interferometer (VLTI) must be know to a sub-millimeter level.

To complicate the fact that the baseline must be known to an extreme precision, the literature defines 
several baselines \citep{1999ApJ...510..505C,2004SPIE.5491.1649H,2013A&A...551A..52S}. From the simple definition of the baseline by \citet{2000plbs.conf.....L} (the distance between two apertures), emerged the idea of a Wide Angle Baseline (WAB), a Narrow Angle Baseline \citep[NAB,][]{2009NewAR..53..344C} and an Imaging Baseline \citep[IMB,][]{PI}. The idea behind these definitions was that each baseline corresponds to a given usage of the interferometer.

 But all these baselines affect the astrometric measurement to various degrees. \citet{PI} used the so-called "limit-points" to define the baselines vectors. Instead of defining the vector by its direction and length, the baseline vector is defined by the point from where it emerges to the point where it goes: the vector is also defined by its geographical position. In this paper, we establish how the position of the various limit-points affect the OPD measurement. Errors due to optical aberrations are intrinsically linked to the position of the limit-points, and we will see that they are an important part of astrometric limitations. Throughout this paper, we will build an error budget, focusing on the biases that occur due to baseline errors and optical aberrations, and showing how GRAVITY, through pupil control, plans to keep these bias below $10\ \mu$as. Note that this paper does not intend to cover random process errors like photon noise, detector noise or background noise.

This paper starts with a presentation of ``The baseline paradox of optical interferometry''. Described in
section~\ref{secparad}, it shows in a pedagogical way how a too simple assumption on the concept of baseline leads to conflicting observations. In
section~\ref{secBaselines}, we use a three dimensional model of the interferometer to calculate the impact of the different baselines limit-points on the OPD measurement. However, this analytical modeling is not enough to account for aberrations of order higher than the tip-tilt. So in section~\ref{secHO} we simplify the problem to two dimension to investigate how these aberrations modify the OPD. The rest of the sections apply those results to the GRAVITY instrument. Section~\ref{secgravity} describes how the instrument works, and the technical choices made for the astrometric measurement. Section~\ref{secerror} puts concrete numbers on the errors and the way they affect astrometry. The end-product of this section is the error budget presented in Table~\ref{table:1}. Section~\ref{secconclusion} concludes.

\section{On the necessity of multiple baselines}
\label{secparad}

\subsection{An apparent paradox}

   \begin{figure}
   \centering
   \resizebox{\hsize}{!}{
   \includegraphics{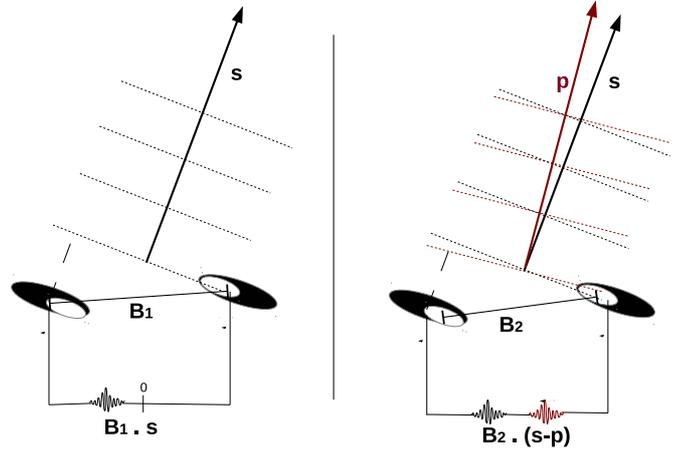}}
      \caption{{\bf The baseline paradox.} Consider a thought experiment where an interferometer is equipped with two masks which rotate in the pupil of
both telescopes. By doing so, the distance between the two apertures changes with time while the telescopes stay firm. In
the left panel, the
observation is done on a single target, acquired on the axis of the telescopes (perpendicularly to the plane of the
aperture). The OPD measurement is done between the white-light fringe and the zero reference of the interferometer. {\em The
OPD does not change with the rotation of the masks}. In the right panel, the interferometer is recording two fringe packets obtained by observing two stars in the same field. The difference in OPD is measured between the two white-light fringes. {\em $\delta$OPD does change with the rotation of the masks}.
              }
         \label{figParadox}
   \end{figure}

Consider a two-telescope interferometer as depicted in Fig.~\ref{figParadox}. In front of each telescope there is an
opaque mask with a circular aperture. The aperture is not centered with respect to the optical axis of the telescope. The
two masks are rotating, moving the respective holes in different directions. 

In the first case (left panel), the interferometer looks at a single star. This star is exactly in the direction of
observation of each telescope. For an unresolved object, the incoming wavefront arrives perpendicularly to the pupil of the
telescope\footnote{we assume here that no atmospheric perturbations are present}. As such, the masks cut out part of the
incoming wavefront. But what matters is the cross-product between the aperture and the phase of the wavefront over the pupil: the wavefront being flat over the pupil, the result of the cross-product does not depend on the position of the hole. In fact, the phase averaged over the aperture is the
same as it would have been obtained without any mask. As a result, the position of the white-light fringe in the beam
combiner is not affected by the rotation of the two masks, even though the distance between the two
apertures changes. The optical path difference between the white-light fringe and the zero of the interferometer depends on the distance between the two telescopes ($\vec {B_1}$):
\begin{equation}
 OPD = \vec s \cdot \vec {B_1}\,.
\end{equation}
The zero of the interferometer is the point at equal optical path distance of the pupils of the two telescopes.

In the second case (right panel of Fig.~\ref{figParadox}), we are not interested in the exact OPD of the fringes, but
in the differential OPD between two fringe patterns. The interferometer is simultaneously observing two celestial
sources of the same field. Eq.~(\ref{eq1}) becomes a differential measurement:
\begin{equation}
 \delta OPD=(\vec s- \vec p) \cdot \vec {B_2}\,
\end{equation}
where $(\vec s- \vec p)$ is the difference in astrometric position between the two sources, and $\delta OPD$ the
difference in position between the two white-light fringes. In this case, the delay in optical length between the fringe
patterns matches the distance between the two apertures, projected in the direction of the source. This time, the relevant
baseline is the one which is modulated ($\vec {B_2}$), and the differential position of the fringes will move when the
masks rotate.

Hence the paradox is that, on two identical interferometers, two behaviors for the fringes positions can co-exist. In one case (single star on-axis),
the fringe pattern is fixed in absolute position. While for the other case (two stars)  one fringe pattern is moving
with respect to the other. In other words, how is it possible that two different baselines $\vec {B_1}$ and $\vec {B_2}$ 
can simultaneously exist and apply to the same interferometer?

\subsection{Solution}

To face this paradox, \citep{PI} proposed several definition of baselines, which would apply to
different kinds of interferometers (or different uses of the same interferometer, as shown above). The idea is that different baselines shall be used depending on the way the OPD is calculated. The wide angle baseline (WAB) is defined by the two limit-points which are the pivot-points of the telescopes (the center of rotation of the telescopes). The imaging baseline (IMB) is defined
by the two limit-points which are the barycentric mean of the effective pupils of each aperture. It must be used when
observing at the same time multiple objects inside a same field of view. Last, the narrow angle baseline (NAB) is
defined by the limit-points which are the image of the metrology end-points in primary space. In the example of the previous section, the $\vec { B_1}$ baseline is the WAB and $\vec { B_2}$ is the IMB.

Note that some baselines are easier to work with than others. The baseline must be {\em stable} to give an accurate astrometric
measurement. An ideal baseline would have two properties:
\begin{enumerate}
 \item It shall be fix with respect to the sidereal reference frame, or at least, fix within a coordinate frame which can be accurately converted into the sidereal frame (e.g. the terrestrial reference frame)
 \item It shall be fix with respect to the interferometric plane where the white-light fringe is observed
\end{enumerate}
The word ``fix'' has the meaning that it does not change with respect to time, telescope pointing nor stellar
object position. Alternatively, instead of being fixed the baseline can be ''fix" can be replaced by "predictable", although it will complicate data reduction. Req.~(1) is to be able to project the baseline in the reference frame of the astronomical object. Req.~(2) is to produce an OPD measurement independent of the position of the objects in the interferometric
field. By definition the pivot-points fulfils Req~(1), while the pupil barycenter fulfils Req (2) \citep{PI} . The NAB does not necessarily fulfil any of those two properties.

\subsection{Practical examples}

Depending on the type of interferometers, some baseline definitions are more useful than others. As an example, the  WAB is null for any kind of single telescope interferometer \citep[like SAM on NACO;][]{2010SPIE.7735E..56T}, as well as an interferometer like the LBT
\citep{2006SPIE.6267E..31H}: $\vec { B_{\rm WAB} }=\vec 0$. This is because the whole structure rotates around a single pivot-point: each entrance aperture has the same pivot point.
For these interferometers, the IMB, which is the distance
between two apertures (LBT), holes (SAM), or mirrors (Michelson stellar interferometer) is the adequate baseline to interpret the data.

Long baseline interferometers with a single beam combiner \citep[for example, AMBER, PIONIER;][]{2007A&A...464....1P,2011A&A...535A..67L}
also require the use of the imaging baseline. However, the imaging baseline is hard to monitor because the effective pupil depends on the
whole optical train, pupil vignetting, telescope pointing, etc... In other words, it does not follow Req (1). So, often and for
convenience, data reduction softwares assume that the IMB is identical to the WAB. However, users of those instruments shall be aware that
it introduces an error in their interferometric measurement.

Finally, in dual feed interferometers like  PTI \citep{1999ApJ...510..505C}, PRIMA \citep{2008NewAR..52..199D}, GRAVITY \citep{2011Msngr.143...16E}, or ASTRA \citep{2010SPIE.7734E..30W}, the metrological link monitors a third baseline often called the NAB. We will show in the following that the NAB should closely match the IMB in order to link the OPD metrology measurement with the angular separation. Moreover, dual feed interferometers have two IMB, one for each combiner. This complicates the equations by adding quasi-static error terms in the OPD measurement.

\section{Baseline limit-points and OPD measurement}
\label{secBaselines}

\subsection{The wide angle baseline}
\label{secWAB}

   \begin{figure}
   \centering
   \includegraphics[width=9cm]{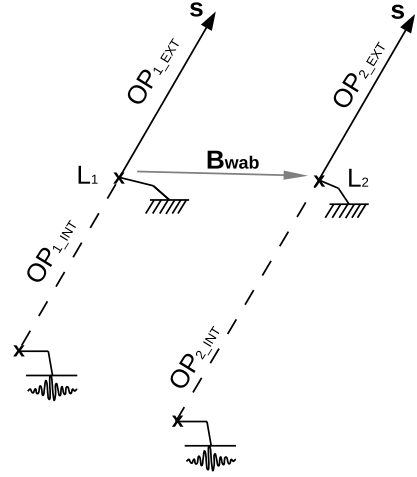}
      \caption{Schematic view of a single beam interferometer. 
      The light paths are represented by light rays. The real light rays are reflected by mirrors up to the beam combiner inside the interferometric lab. However, in this simplified representation, all the mirrors are removed. Thus, the represented virtual light rays are straight lines from the star to the virtual images of the beam combiner. The mechanical structure of the interferometer is reduced to two pivot-points, $\vec {L_1}$ and $\vec {L_2}$ which are the pivot-points of the telescopes. The wide angle
baseline (WAB) is represented as a vector between the two pivot-points: $\vec {B_{\rm WAB}} = \protect \overrightarrow{\vec  { L_1L_2}}$. The OPD measurement is made between the beam combiners and the metrology end-point. It is related to the stellar position through Eq.~(\ref{OPDWAB}).
              }
         \label{figBase1}
   \end{figure}

In Fig.~\ref{figBase1} is the simplest representation possible of a purely theoretical astrometric
interferometer. The telescopes are represented by two  geographically-localized points $\vec {L_1}$ and $\vec {L_2}$. The
telescopes point by rotating around themselves, ie, $\vec {L_1}$ and $\vec {L_2}$ are also the pivot points. All mirrors have been
suppressed in
the representation, and the optical trajectories are represented by virtual rays. Because of this, the beam combiner have two virtual images,
one in each arm of the interferometer (in reality, thanks to folding mirrors, the fringes are -- of course -- physically located
at the same position).

The interferometer metrology probes two optical length measurements, one for each telescope. The astrometric measurement uses the difference between the two:
\begin{equation}
OPD = OP_{2-INT} - OP_{1-INT} \, .
 \label{OPDint}
\end{equation}
The two values $OP_{1-INT}$ and $OP_{2-INT}$ are measured, to a constant, between the {\em central fringe}\footnote{In the absence of dispersion, the central fringe is also the white light fringe, i.e. fringe with maximum contrast.}, and the telescope pivot points $\vec {L_1}$ or $\vec {L_2}$. In the whole Section~\ref{secBaselines}, we will assume that the metrology measurement is obtained with respect to the central fringe, i.e. without any additional OPD introduced at the beam combiner level.

 Since by definition the central fringe is the fringe at equal optical path, we can establish by the following equation the relation between $OP_{INT}$ and $OP_{EXT}$:
\begin{equation}
 OP_{1-EXT} + OP_{1-INT} =   OP_{2-EXT} + OP_{2-INT} \, . 
 \label{OPDequal}
\end{equation}
where the two numbers $OP_{1-EXT}$ and $OP_{2-EXT}$ are the optical path lengths up to
the stellar emission, outside the interferometer.

Last, the equation which links the two external optical paths is the projection in
the stellar direction of the 3D vector delimited by the two pivot points. According to the baseline definitions
in \citet{PI}, this vector is the WAB:
\begin{equation}
 OP_{1-EXT} - OP_{2-EXT} = \vec  s \cdot  \overrightarrow{ \vec { L_1L_2}}
 \label{OPDproj}
\end{equation}
Note that in the whole paper, we will use an arrow to represent a vector which goes from a point $\vec X$ to a point $\vec Y$:
 $\overrightarrow{ \vec {XY } }= \vec { Y} - \vec { X }$.
Here: \begin{equation} \overrightarrow{ \vec { L_1L_2 } }= \vec { L_2 } - \vec { L_1 }= \vec {B_{\rm WAB}}\ .\end{equation}
The advantage of the WAB is that it is theoretically fix with respect to the ground: it does not change with the pointing direction of the telescope.

Eqs.~(\ref{OPDint}),~(\ref{OPDequal}), and~(\ref{OPDproj}) directly lead to the relation which links the OPD
measurement to the stellar position:
\begin{equation}
 OPD = \vec { s } \cdot \vec { B_{\rm WAB}} \ . 
 \label{OPDWAB}
\end{equation}

\subsection{The narrow angle baseline}
\label{secNAB}

   \begin{figure}
   \centering
   \includegraphics[width=8.7cm]{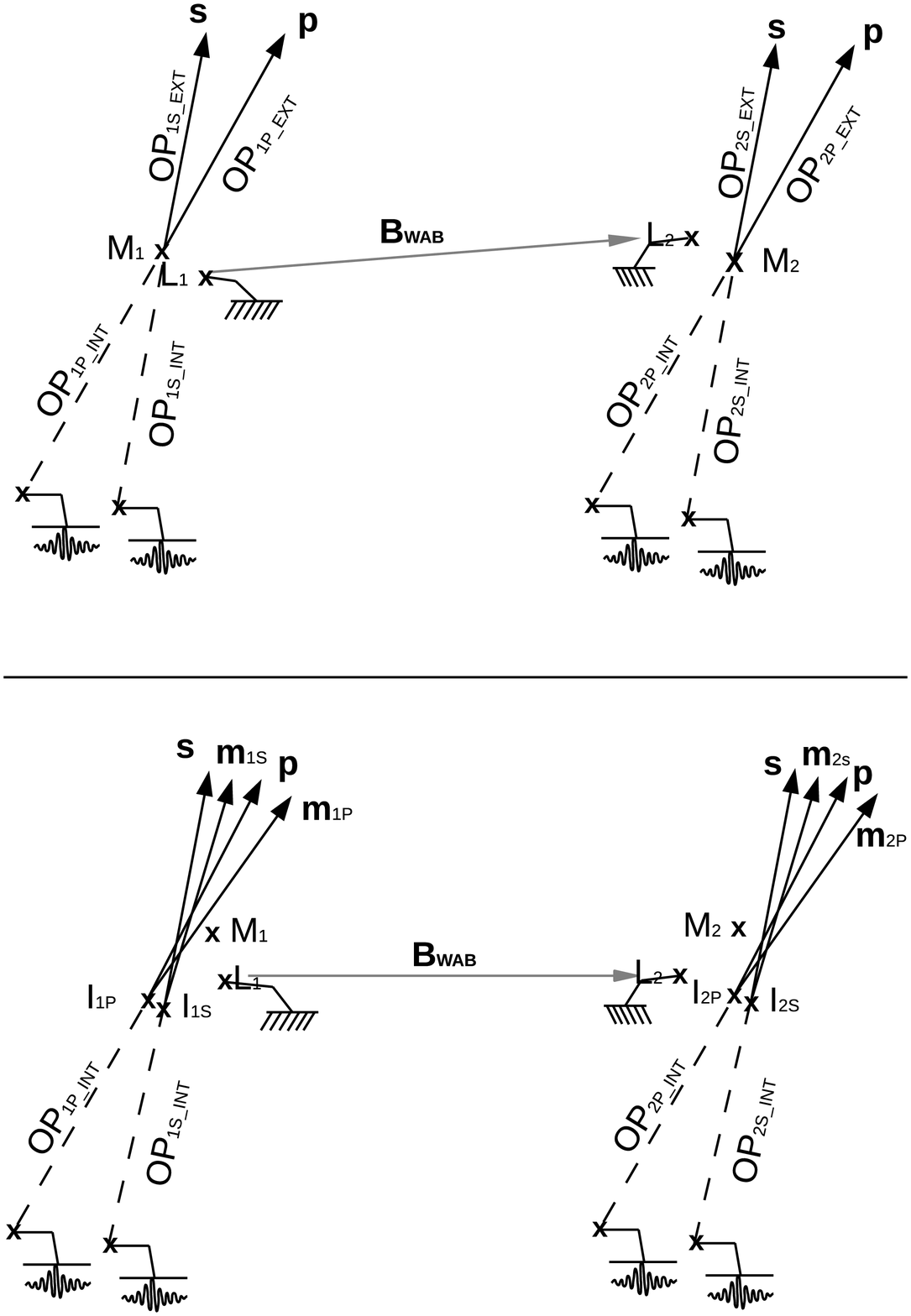}
      \caption{Schematic view of an astrometric interferometer. 
      The light paths are virtual, as seen from the star; all
mirrors are removed, resulting in a straight line from the star to the beam combiner. Only the wide angle
baseline (WAB) is represented, but all the baselines limit-points are represented.
       $\vec {L_1}$ and $\vec {L_2}$ are the pivot-points of the telescopes: $\vec {B_{\rm WAB}} = \protect \overrightarrow{\vec {L_1L_2}}$.
       The $\vec {M_1}$ and $\vec {M_2}$ points are the end-points of the metrology: $\vec {B_{\rm NAB}} =\protect \overrightarrow{\vec { M_1M_2}}$.
       Last, $\vec {I_{P1}}$, $\vec {I_{P2}}$, $\vec {I_{S1}}$  and $\vec {I_{S2}}$ are the barycenters of the effective pupils of each beam combiners: $\vec {B_{\rm IMB-S}} = \protect \overrightarrow{\vec {I_{1S}I_{2S}}}$ for the first combiner and $\vec {B_{\rm IMB-P}} =\protect \overrightarrow{\vec{ I_{1P}I_{2P}}}$ for the second. {\em Upper panel:}   the metrology measurement is made between the beam combiners and the metrology end-point. The end-points are
 different from the pivot-points, which justify in Eq.~(\ref{OPDNAB}) the use of the narrow angle baseline instead of the WAB. {\em
Lower panel:} Full schematic representation of the interferometer, including the presence of the imaging limit points. Imaging limit points are necessary to account for error terms
due to difference in the propagation direction of the metrology ($\vec {m_{1S}}$, $\vec {m_{2S}}$, $\vec {m_{1P}}$ and $\vec {m_{2P}}$) with respect to the stellar beams ($\vec s$ and $\vec p$). It introduces additional error terms
which are stated in Eq.~(\ref{OPDIMG}).
              }
         \label{figBase2}
   \end{figure}

The upper panel of Fig.~\ref{figBase2} increases the complexity with respect to Fig.~\ref{figBase1}  by including four measurements instead of two:
the interferometer
is now observing two different stars: $\vec p$ and $\vec s$. The OP measurements are made between the white-light fringes and the
end-points of the metrology signal, defined in geographic space: $\vec {M_1}$ and $\vec {M_2}$. Note that the end-points can be
inside the optical train. If this is the case,  $\vec {M_1}$ and $\vec {M_2}$ are the virtual images of the metrology end-points, as
seen from the star (in the sidereal reference frame).
Then, the OPD measurement is twice differential:
\begin{equation}
\delta OPD = OP_{2S-INT} - OP_{1S-INT} +  OP_{1P-INT} - OP_{2P-INT}\, . 
\label{eqOPDm}
\end{equation}

Similarly to section~\ref{secWAB}, we can combine Eq.~(\ref{OPDequal}) -- which links the internal OPs to the external
OPs -- to Eq.~(\ref{OPDproj}) -- which links the external OPs to the baseline vector $ \overrightarrow{ \vec { M_2 M_1} }$ -- to derive:
\begin{eqnarray}
 OP_{2S-INT} - OP_{1S-INT} = \vec  s \cdot \overrightarrow{\vec { M_1M_2}} \label{OPDMM1}& \\
 OP_{2P-INT} - OP_{1P-INT} = \vec  p \cdot \overrightarrow{\vec { M_1M_2}} \label{OPDMM2}&.
\end{eqnarray}
Then, we can define the narrow angle baseline: 
\begin{equation}  \overrightarrow{\vec { M_1M_2}} = \overrightarrow{ \vec { M_1L_1} }+  \vec { B_{\rm WAB} }+ \overrightarrow{\vec { L_2M_2}}=\vec { B_{\rm NAB}}\ .\end{equation}
  As shown in \citet{PI} it
is the one which must be used to derive the separation between the two stellar objects:
\begin{equation}
\delta OPD = (\vec s - \vec p) \cdot \vec { B_{\rm NAB}}\,.
 \label{OPDNAB}
\end{equation}

 A difficulty of the narrow angle baseline is that the limit-points (the end-points of the metrology) are generally defined in the telescope reference frame. As a consequence, the NAB depends on the telescope pointing. Also, if the metrology end-points are not physically located in the telescope frame, but inside the intereferometric optical train, the narrow angle baseline is more difficult to calculate. It is then defined by the virtual images of the metrology end-point, as seen from outside the telescope.

\subsection{The imaging baseline}
\label{secIMB}

It is necessary to introduce the notion of the imaging baseline once we want to consider the impact of field position of the targets with respect to the metrology. It can be traced back to the van-Cittert Zernike theorem, which links contrast and phase to the spatial geometry of the astrophysical object \citep{PI}.

For a dual beam interferometer like GRAVITY, there are two imaging baselines: one for each beam combiner. This is due to the fact that the limit-points of the imaging baseline is defined by the effective pupil, which is different from the telescope pupil. For a single-mode interferometer the effective pupil can be properly defined. It is the acceptance mode of the single-mode fibers as seen from the star (viewed through the telescope pupil). For a dual beam combiner there are four acceptance modes, thus four limit-points or two IMBs. In the lower panel of Fig.~\ref{figBase2}, the limit-points are noted $\vec {I_{1S}}$ and  $\vec {I_{2S}}$  for the first beam combiner, and $\vec {I_{1P}}$ and  $\vec {I_{2P}}$ for the second.
These limit-points do not have to be co-located with the telescope pupil center: this is called a pupil positioning error,
which can be both lateral or longitudinal.

The propagation directions of the metrology beams are represented in Fig.~\ref{figBase2} by the vectors $\vec { m_{1S}}$, $\vec { m_{2S}}$, $\vec { m_{1P}}$ and $\vec { m_{2P}}$. The propagation directions of the incoming light of the stellar sources are still $\vec s$ and $\vec  p$. Hence, for example, the vector difference $(\vec { m_{2S} }- \vec  s)$ characterizes the pointing error of the metrology of the first beam combiner on the second telescope toward the source denoted $s$. This is equivalent to a tip-tilt error.

The  imaging baseline limit-points have, by definition, an important property: the optical path length between the fringes and the limit-points is the same for the metrology and for the stellar light. This is a property of the effective pupil: the pupil is field-independent\footnote{The demonstration for this is in the appendix of \citet{PI}. {\em Basically}: from the imaging baseline perspective, any angular misalignment (star light versus metrology path co-aligned with fringe sensor lobe) is interpretable as an external mis-alignment, leaving the internal metrology versus star light optical path identical.
}. As a consequence, $OP_{1S-INT}$ defined between the beam combiner and the imaging baseline limit-point
does not depend on the  $(\vec { m_{1S} }- \vec  s)$ tilt error. This is also the case for the three other optical path lengths: $OP_{2S-INT}$, $OP_{1P-INT}$ and $OP_{2P-INT}$.

However, the $\delta OPD$ measurement is done up to the metrology end-point. Following the conventions from the lower panel of Fig.~\ref{figBase2}, Eq~(\ref{eqOPDm}) must be rewritten to include the optical path length corresponding to the projected distance between the imaging and metrology end-points (the four $\vec  m \cdot \vec { IM}$ terms). Thus, $\delta OPD$ measured between $\vec{M_1}$ and $\vec{M_2}$ becomes:
\begin{eqnarray}
\delta OPD_{M_{12}} &=& OP_{2S-INT}  +  \vec { m_{2S}} \cdot  \overrightarrow{ \vec { I_{2S}M_2}} \nonumber  \\
&&- OP_{1S-INT} -  \vec { m_{1S}} \cdot \overrightarrow{ \vec { I_{1S}M_1}} \nonumber  \\
&&+  OP_{1P-INT} +  \vec { m_{1P}} \cdot \overrightarrow{ \vec { I_{1P}M_1}} \nonumber  \\
&&- OP_{2P-INT} -  \vec { m_{2P}} \cdot  \overrightarrow{ \vec { I_{2P}M_2}} \, .  \label{OPDII0}
\end{eqnarray}
On the other hand, the internal and external optical path lengths must satisfy Eq.~(\ref{OPDequal}). In the same way we
obtained Eq.~(\ref{OPDMM1}) and Eq.~(\ref{OPDMM2}), we can derive:
\begin{eqnarray}
 OP_{2S-INT} - OP_{1S-INT} &=& \vec  s \cdot \overrightarrow{ \vec {I_{1S}I_{2S}}}  \label{OPDII1} \\
 OP_{2P-INT} - OP_{1P-INT} &=& \vec  p \cdot \overrightarrow{ \vec {I_{1S}I_{2S}}}\label{OPDII2} \ .
\end{eqnarray}
where the vectors $\overrightarrow{ \vec {I_{1S}I_{2S}}}$ and $\overrightarrow{ \vec {I_{1S}I_{2S}}}$ correspond to the two imaging baselines:
\begin{eqnarray}
\overrightarrow{ \vec {I_{1S}I_{2S}}}&=&\overrightarrow{ \vec {I_{1S}L_1}} +\vec  { B_{\rm WAB}} + \overrightarrow{ \vec  {L_2I_{2S}}}= \vec { B_{\rm IMB-S}} \label{OPDII3} \\
\overrightarrow{ \vec {I_{1P}I_{2P}}}&=&  \overrightarrow{ \vec {I_{1P}L_1}} +\vec  { B_{\rm WAB}} + \overrightarrow{ \vec  {L_2I_{2P}}}=  \vec { B_{\rm IMB-P}} \label{OPDII4} \ .
\end{eqnarray} 

Combining together Eqs.~(\ref{OPDII0}), (\ref{OPDII1}), (\ref{OPDII2}), (\ref{OPDII3}) and (\ref{OPDII4}), we can
link the $\delta OPD_{M_{12}}$ measurement to the differential angular position $(\vec s - \vec  p)$ and to the wide angle baseline $\vec  { B_{\rm WAB}}$:
\begin{eqnarray}
\delta OPD_{M_{12}} &=& (\vec  s - \vec  p) \cdot \vec { B_{\rm WAB} }\nonumber \\
 &&+ \vec { m_{2S}} \cdot \overrightarrow{ \vec { I_{2S}M_2}} -  \vec { m_{1S} }\cdot \overrightarrow{ \vec {I_{1S}M_1}} \nonumber    \\
 && +  \vec { m_{1P}} \cdot \overrightarrow{ \vec { I_{1P}M_1}} -  \vec { m_{2P}} \cdot \overrightarrow{ \vec {I_{2P}M_2}} \nonumber  \\
 && + \vec s \cdot (\overrightarrow{ \vec { I_{1S}L_1}} + \overrightarrow{ \vec {I_{2S}L_2}}) \nonumber  \\
 && - \vec p \cdot (\overrightarrow{ \vec { I_{1P}L_1}} + \overrightarrow{ \vec {I_{2P}L_2}}) \,.
\end{eqnarray}
This equation simplifies to:
\begin{eqnarray}
\delta  OPD_{M_{12}} &=& (\vec  s - \vec  p) \cdot (\vec { B_{\rm WAB}} + \overrightarrow{\vec {M_1L_1}} - \overrightarrow{\vec {M_2L_2}})  \nonumber \\
 &&+ (\vec { m_{2S}} - \vec  s) \cdot \overrightarrow{\vec {I_{2S}M_2}} -  (\vec {m_{1S}} - \vec s) \cdot \overrightarrow{\vec {I_{1S}M_1}   } \nonumber \\
 && +  (\vec {m_{1P}} - \vec  p) \cdot \overrightarrow{\vec {I_{1P}M_1}} -  (\vec {m_{2P}} - \vec p) \cdot \overrightarrow{\vec {I_{2P}M_2}} \,.
 \label{OPDIMGL}
\end{eqnarray}
or
\begin{eqnarray}
\delta  OPD_{M_{12}} &=& (\vec s - \vec p) \cdot \vec { B_{\rm NAB}} \nonumber \\
 &&+ (\vec { m_{2S}} - \vec  s) \cdot \overrightarrow{\vec {I_{2S}M_2}} -  (\vec {m_{1S}} - \vec s) \cdot \overrightarrow{\vec {I_{1S}M_1}}    \nonumber \\
 && +  (\vec { m_{1P}} - \vec p) \cdot \overrightarrow{\vec {I_{1P}M_1}} -  (\vec {m_{2P}} - \vec p) \cdot \overrightarrow{\vec { I_{2P}M_2}} \,.
 \label{OPDIMG}
\end{eqnarray}

Hence,  $\delta OPD_{M_{12}}$ results from the sum of two components. The first component, $(\vec s - \vec p) \cdot \vec { B_{\rm NAB}}$,
is the main contributor to the full measurement.
{\em It reflects the assertion stated in \citet{PI} that
the NAB baseline is the main baseline of a dual feed astrometric interferometer}. 
However several second-order terms are present due to that fact that the imaging baseline 
limit-points are not necessarily the same as the
metrology end-points ($\overrightarrow{\vec {IM}} \neq \overrightarrow{ \vec 0}$). Since both $\overrightarrow{\vec {IM}}$ and $(\vec  m - \vec s)$ are small, the amplitude of this component will be small with regard to the total optical path difference. However,
if no care is taken, it adds to the $\delta OPD_{M_{12}}$ in a way that it will bias the estimation of the angular separation $(\vec  s -
\vec  p)$.

Note that it is possible to combine the second-order terms together to show the differential between the imaging baselines $\vec { B_{\rm IMB-S}} $ and $\vec { B_{\rm IMB-P}} $ and the narrow angle baseline $\vec { B_{\rm NAB}} $  :
\begin{eqnarray}
\delta  OPD_{M_{12}} &=& (\vec s - \vec p) \cdot \vec { B_{\rm NAB}} \nonumber \\
 &&+ (\frac{\vec { m_{2S}}+\vec { m_{1S}}}{2} - \vec  s) \cdot ( \vec { B_{\rm NAB}} -  \vec { B_{\rm IMB-S}}) \nonumber \\
 &&-  (\frac{\vec { m_{2P}}+\vec { m_{1P}}}{2} - \vec  p) \cdot ( \vec { B_{\rm NAB}} -  \vec { B_{\rm IMB-P}}) \nonumber \\
 &&+  (\frac{\vec { m_{2S}}-\vec { m_{1S}}}{4} + \frac{\vec { m_{2P}}-\vec { m_{1P}}}{4} ) \cdot ( \overrightarrow{\vec {  I_{1S}I_{1P}}} + \overrightarrow{ \vec { I_{2S}I_{2P}}}) \nonumber \\
 &&+  (\frac{\vec { m_{2S}}-\vec { m_{2P}}}{4} - \frac{\vec { m_{1S}}-\vec { m_{1P}}}{4} ) \cdot (\overrightarrow{\vec { I_{1S}M_1}} +\overrightarrow{ \vec { I_{1P}M_1}})\nonumber \\
 &&+  (\frac{\vec { m_{2S}}-\vec { m_{2P}}}{4} - \frac{\vec { m_{1S}}-\vec { m_{1P}}}{4} ) \cdot (\overrightarrow{\vec { I_{2S}M_2}} +\overrightarrow{\vec { I_{2P}M_2}})   \,.\nonumber \\
 \label{OPDIMG123}
\end{eqnarray}
According to this formula, $\delta  OPD_{M_{12}}$ can be equal to  $\delta  OPD$ as spelled in Eq~(\ref{OPDNAB}) if all the three following conditions are fulfilled:
\begin{enumerate}
\item The baselines match each other: $ \vec { B_{\rm NAB}} =  \vec { B_{\rm IMB-S}}=  \vec { B_{\rm IMB-P}}$.
\item The limit-points of the two imaging baselines are superposed: $ \vec {  I_{1P}} - \vec I_{1S} =  \vec  I_{2S} - \vec I_{2P}$ (equal to $\vec 0$ asuming 1.).
\item The metrology beams have the same  propagation angle on each telescope: ${\vec { m_{1S}}-\vec { m_{1P}}} = {\vec { m_{2S}}-\vec { m_{2P}}} $.
\end{enumerate}
 This is {\em not} a situation that we will assume in 
the following GRAVITY error budget.

\section{High-order optical aberrations}
\label{secHO}

The notion of imaging baseline is intrinsically related to the presence of a pointing error, or tip-tilt ($\vec m - \vec s \neq \vec 0$). These tilt-tilt errors are equivalent to low order optical aberrations, and are considered in the previous section. In this section, we  include the effect of the higher order Zernike polynomials.

To be exact, a full diffraction analysis would have to be performed over the whole metrology train. That would mean including a Fresnel analysis over the optics down to the fiber for the stellar light, and up to the metrology sensor for the metrology beam. However, we can considerably simplify the problem by neglecting the effect of the $z$ direction inside the optical train. It is equivalent to consider that the atmosphere, the metrology sensor, the pupil of the telescopes, the pupil of the instruments and the optical aberrations are conjugated inside a single plane. In that plane, only the two lateral directions ($x$ and $y$) matter. Then, the effect of the high-order aberrations can be seen as a problem of a two-dimensional coupling of the light onto the modes of the beam combiners.

\subsection{Aberrations affecting the stellar light}

In the case of a single-mode interferometer, the modes of the beam combiners can be identified as the modes of acceptance of the single-mode fibers. The phase delays due to the optical aberrations are therefore the argument of the coupling coefficient of the abnormal light onto the mode of each fiber: $\delta \Phi= \arg( \rho)$, where $\delta \Phi$ is the phase error caused by the aberration, and $ \rho$ the coupling coefficient of the light into the fiber \citep{1959ITAP....7..118R,1988ApOpt..27.2334S}:
\begin{equation}
\rho = \frac{\iint \limits_{pupil}  \!   E(x,y)    E_{F}^*(x,y) \,  \mathrm dx \mathrm dy }{\sqrt{\iint \limits_{pupil}  \!   |E(x,y)_s|^2 \,  \mathrm dx \mathrm dy \iint \limits_{pupil}  \!   |E_{F}(x,y)|^2 \,  \mathrm dx \mathrm dy}}
\end{equation}
where $E(x,y)$ is the incoming aberrated wavefront and $E_{F}(x,y)$ is the acceptance mode for the electric field of the single-mode fiber. The cross-correlation is easier to understand in the focal plane, i.e. at the entrance of the fiber. However, according to the Parseval-Plancherel theorem, it can also be calculated in the pupil plane, which turns out to be more convenient to model the aberrations. 

To calculate the impact on the OPD measurement, we have to consider the modes of all four fibers ($E_{Fs1}$, $E_{Fs2}$, $E_{Fp1}$ and $E_{Fp2}$) and of the four distinct incoming  wavefront ($E_{s1}$, $E_{s2}$, $E_{p1}$ and $E_{p2}$). Then, the phase error on the OPD measurement will be:
\begin{eqnarray}
\delta \Phi_{\rm stellar}&=&\arg \left(  \iint \limits_{\rm pupil}  \!    E_{s1}(x,y)    E^*_{Fs1}(x,y) \,  \mathrm dx \mathrm dy     \right) \nonumber \\
	&&-\arg \left(  \iint \limits_{\rm pupil}  \!    E_{s2}(x,y)    E^*_{Fs2}(x,y) \,  \mathrm dx \mathrm dy     \right) \nonumber \\
&&-\arg \left(  \iint \limits_{\rm pupil}  \!    E_{p1}(x,y)    E^*_{Fp1}(x,y) \,  \mathrm dx \mathrm dy     \right) \nonumber \\
&&+\arg \left(  \iint \limits_{\rm pupil}  \!    E_{p2}(x,y)    E^*_{Fp2}(x,y) \,  \mathrm dx \mathrm dy     \right) \label{eq:phistel}
\end{eqnarray}
where the integration is done over the electric fields present inside the pupil.

\subsection{Aberrations affecting the metrology light}

While the phases measured inside the beam combiners are affected by the full aberration of the stellar light, the metrology value is solely modified by the optical aberrations inside the optical train. These aberrations can be quantified according to the Zernike polynomials $Z^n(x,y)$ \citep[$n$ according to numbering by][]{1976JOSA...66..207N} of variance unity. Each one of the four metrology beams can be affected by these aberrations. However, only a differential aberration will affect the differential metrology measurement. This is a beam-walk effect: the two stellar beams have different footprints on the optics, which can result in a differential optical aberration. Assuming that the aberrations are a sum of Zernike of amplitude $k^n_{s1}$, $k^n_{s2}$, $k^n_{p1}$ and $k^n_{p2}$ (one for each one of the four metrology beams), we can derive a phase error to the metrology:
\begin{eqnarray}
\delta \Phi_{\rm met}&=&\iint \limits_{\rm receivers\ tel1}  \!  \sum_n \left[ (k^n_{s1} - k^n_{p1})\ Z^n(x,y) \right] \mathrm dx \mathrm dy     \nonumber \\
	&&-\iint \limits_{\rm receivers\ tel2}  \!  \sum_n \left[  (k^n_{s2} - k^n_{p2})\ Z^n(x,y)  \right] \mathrm dx \mathrm dy\label{eq:phimet}
\end{eqnarray}
The full metrology measurement is a phase measurement averaged over all the metrology receivers. This is a general case. Depending on the type of metrology system, and position/number of receivers, this integral can be a simple summation of several measurements at given $(x,y)$ positions.

\section{The GRAVITY instrument}
\label{secgravity}

\subsection{GRAVITY metrology principle}
\label{design}

   \begin{figure*}
   \centering
  \resizebox{\hsize}{!}{
   \includegraphics{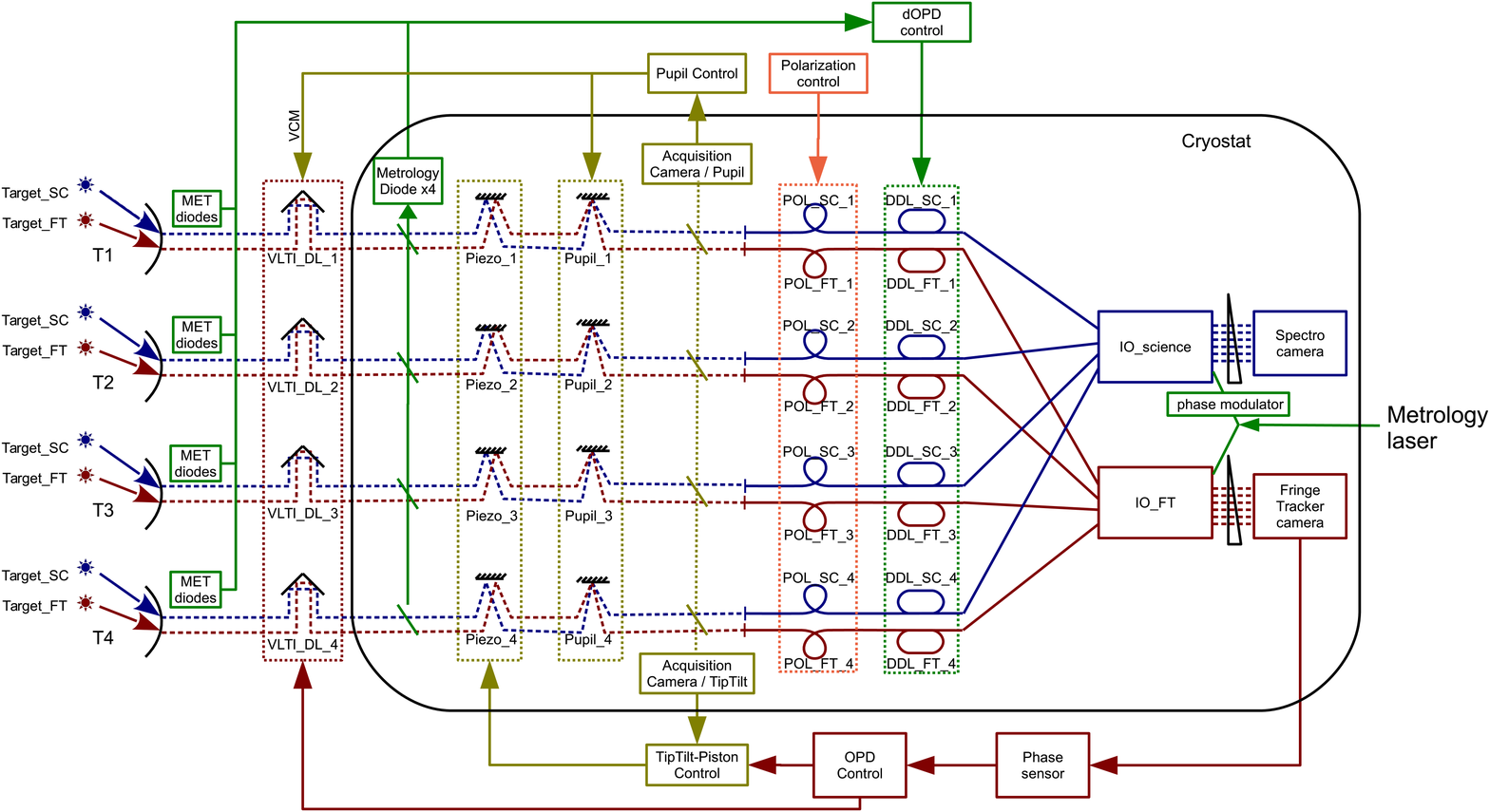}}
      \caption{ The GRAVITY active subsystems involving the astrometric measurement. In red is the optical path of the reference star. In blue the optical path of the science star. The dotted lines are free space light propagation, and the solid lines correspond to the single-mode fibers.  The reference star is used to correct in real time the atmospheric piston by activating the internal piezo and/or the VLTI delay lines. In green is represented the metrology system, injected into the two integrated optics beam combiners (IO), and measured thanks to 4 metrology diodes inside the cryostat, and 4 metrology diodes on each telescope. The metrology measurements are used to control the GRAVITY differential delay lines to position the fringes on the science camera. In yellow is represented the acquisition camera control loop which commands the tip-tilt  and the pupil position. In orange is the open loop polarization control system.
              }
         \label{figGRAVMet}
   \end{figure*}
   
As mentioned before, GRAVITY is a 4-telescope interferometer for the VLTI. It is operating in the $K$ band and is able to couple the four 8.2\,m  Unit Telescopes (UTs) or the four 1.8\,m Auxiliary Telescopes (ATs). It is assisted with a visible and infrared adaptive optics system for the UTs. GRAVITY is doted with two beam combiners in integrated optics \citep{2012SPIE.8445E..2XJ}, one for a reference source and the other for the science target. In this dual mode, the goal of GRAVITY is to provide an accurate differential phase measurement to achieve astrometry with a 10 micro-arcsecond accuracy in a 2-arcsecond field (4-arcsecond field with the ATs).

The reference source is used to cancel atmospheric perturbations over the 6-baselines thanks to a dedicated fringe tracking system. To do so, the GRAVITY instrument will have a fast feedback control loop using steering mirrors on piezo actuators inside the cryostat (with a cutoff bandwidth close to 30Hz), and a slow feedback (for offloading) to the main VLTI delay lines. GRAVITY will have to fringe track on a K=10 unresolved object with 300 nm OPD residuals, setting the basic specification of the system. 

The basic functions impending the OPD measurement are presented in Fig.~\ref{figGRAVMet}. The metrology is injected into both beam combiners \citep{2012SPIE.8445E..1OG}. On one beam combiner, the metrology is phase-modulated with respect to the other with a Lithium Niobate phase modulator. Going through the optics in reverse direction of the stellar light, the metrology beams encounter respectively in the cryostat \citep{2012SPIE.8445E..2VH}: the fibered differential delay lines, the fibered polarization rotators, the pupil actuator mirror, and the fast piezo actuators. Before exiting the cryostat, part of the metrology light is recombined on internal metrology diodes to provide a constant and reliable feedback for the differential fiber delay lines (without risk of beam interruption). The rest of the metrology light is back propagated through the VLTI optical train, across all the optics including M1, up to receivers situated on each four spider arms of each four telescopes. For each telescope a phase value is extracted from the mean of the phase value of the four telescope diodes: $\Phi_{Mi}$, with $i$ the telescope number.

The stellar light is following a similar path in reverse. Once in the cryostat, after the pupil actuator mirror and before injection into the fluoride fibers, the astrophysical sources are separated with a roof mirror acting as a field separator 
\citep{2012SPIE.8445E..1UP}. The light is guided inside the single-mode fibers up to integrated optics beam combiners, the outputs of which are dispersed and imaged on infrared cameras. On one hand, the light of the bright reference source is send to a fast eAPD detector \citep[The SAPHIRA detector,][]{2012SPIE.8453E..0TF} from which is read a phase measurement for each 5 wavelength channels of the fringe tracker: $\Phi_{P}(\lambda_s)$. This phase is kept close to zero by the means of the fringe tracker commanding the piezo actuators. On the other hand, the light of the science target is send to a Hawaii 2 which is slower (DIT time of the order of the second) but larger (2k by 2k pixels) than the SAPHIRA detector. On the Hawaii 2, the light is spectrally dispersed giving a resolving power of $\Delta \lambda/\lambda = 20$, 500 or 4\,000 \citep{2012SPIE.8445E..2RS}. On this detector one obtains the phase $\Phi_{S}(\lambda_s)$. For simplicity, we will consider phase measurements on both detectors to be done at a common wavelength $\lambda_s$.

All these phase measurements are combined to give 6 differential OPD measurements: one for each baseline. For example, the optical delay $\delta  OPD_{M_{12}}$ is obtained between telescope 1 and 2 by the relation:
\begin{equation}
\delta  OPD_{M_{12}} = (\Phi_{S} - \Phi_{P}) \frac {\lambda_s}{2 \pi} + (\Phi_{M_2}-\Phi_{M_1})\frac{\lambda_{m}}{2 \pi} 
\label{eqOPDPhi}
\end{equation}

\subsection{Dispersion in GRAVITY}

In Eq.~(\ref{eqOPDPhi}) is missing an important term caused by dispersion. Dispersion can have an important impact on the contrast of the white-light fringes \citep{1995A&A...293..278C}. But it also have an important effect on the astrometry. It comes from the fact that the apparent optical path length depends on the wavelength, as the OP length is proportional to the index of refraction of the medium through which it propagates. It writes:
\begin{equation}
OP(\lambda)=n(\lambda)\cdot L
\end{equation}
where $OP(\lambda)$ is the apparent optical path length and $L$ the physical length.
For an addition of several media along the optical path length measured by the metrology,  dispersion introduces an OPD error equal to:
\begin{equation} 
D(\lambda_{m},\lambda_s)=\sum_{med} \Delta_{med} \cdot(n^{\lambda_m}_{med}-n^{\lambda_s}_{med}) 
\end{equation}
where $\Delta_{med}$ is the difference of physical length of each optical medium, and $n^{\lambda}_{med}$ is the refractive index of the corresponding medium at a  wavelength $\lambda$.

In GRAVITY, the wavelength of the metrology ($\lambda_{m}=1.908 \mu$m) is different from the wavelength of the stellar light ($1.95\mu m < \lambda_s < 2.45 \mu m$). On the path of the metrology, there are 3 different media: air, vacuum (inside the cryostat), and fluoride glass (fibers). The dispersion of vacuum is null. So the dispersion term can be written:
\begin{equation}
 D(\lambda_{m},\lambda_s)= \Delta_{air} \cdot(n^{\lambda_m}_{air}-n^{\lambda_s}_{air}) + \Delta_{fiber} \cdot (n^{\lambda_m}_{fiber}-n^{\lambda_s}_{fiber})\,.
\label{eqDisp}
\end{equation}

Combining Eq.~(\ref{eqOPDPhi}) with Eq.~(\ref{OPDIMG}), and including the effect of dispersion, we can now write down the analytical formula which link the phase measurements to the angular separation of the targets on sky :
\begin{eqnarray}
(\vec  s - \vec  p) \cdot \vec { B_{\rm NAB} } &=&(\Phi_{S}  - \Phi_{P})\frac{\lambda_{s}}{2 \pi} +(\Phi_{M_2}-\Phi_{M_1})\frac{\lambda_{m}}{2 \pi}  -  D(\lambda_{m},\lambda_s) \nonumber \\
&&+ {\rm error\ terms\ \ (2^{nd}\,order)}\,
\label{eqOPDtotal}
\end{eqnarray}
It is not foreseen to calculate the value of the second order error terms. The goal is to keep them low enough in the error budget.
They come from baseline errors and optical aberrations, as spelled in Eq.~(\ref{OPDIMG}), Eq.~(\ref{eq:phistel}) and Eq.~(\ref{eq:phimet}):
\begin{eqnarray}
 {\rm error\ terms}&=& (\vec { m_{1S}} - \vec  s) \cdot\overrightarrow{ \vec {I_{1S}M_1}} -  (\vec { m_{2S}} - \vec  s)  \cdot \overrightarrow{\vec {I_{2S}M_2}}    \nonumber \\
 &&  + (\vec { m_{2P}} - \vec  p) \cdot \overrightarrow{\vec {I_{2P}M_2}} - ( \vec { m_{1P}} - \vec p) \cdot \overrightarrow{\vec {I_{1P}M_1}} \,. \nonumber \\
 &&+ \delta \Phi_{\rm stellar}\frac{\lambda_{s}}{2 \pi} +\delta \Phi_{\rm met}\frac{\lambda_{m}}{2 \pi}
\label{eqOPDerr}
\end{eqnarray}

\subsection{Geographical position of the limit-points of GRAVITY}
\label{BaseSpec}

   \begin{figure*}
\centering
   \includegraphics[width=11cm]{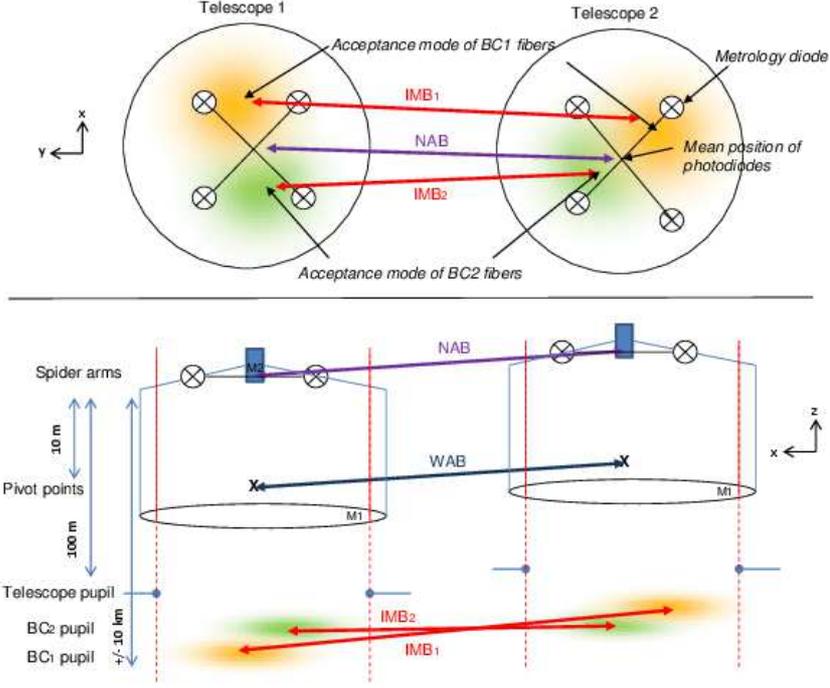}
      \caption{Baseline representation for a two-telescopes interferometer according to the GRAVITY concept. The
baselines are defined by their limit-points. The WAB is defined by the two pivot-points (which defines the center of the
telescopes reference frames). The NAB is defined by the geometrical average of the position of the four metrology
photodiodes situated on the spider arms. The coordinate of the NAB limit points are ($x$,$y$,$z$) = (0,0,10\ m). The
two imaging baselines IMB$_1$ and
IMB$_2$ (one for each beam combiner) have limit-points which correspond to the pupil position of their respective beam
combiners, weighted laterally by the telescope aperture as well as the electric-field acceptance of the single mode
fibers. Internal pupil tracking inside the GRAVITY instrument will track the spider arms to place the IMB limit points
on the NAB limit point. Expected lateral position will be $|x,y| = 0 \pm 4$\ cm. Expected longitudinal position will be $z
= 0 \pm 10$\ km.
              }
         \label{figGRAV}
   \end{figure*}

The idea that drove the design of the GRAVITY instrument is therefore to have monitoring and steering possibilities for both the pupil and the field plane. This is paramount to keep under control the error terms in Eq.~(\ref{eqOPDerr}). Focal and pupil plane positions are simultaneously supervised during observation by a dedicated camera situated inside the cryostat \citep{2012SPIE.8445E..34A}. 

The feedback system of this camera -- called the acquisition camera -- is presented in yellow in Fig.~\ref{figGRAVMet}. The first terms to minimize are the field terms: $\vec { m_{iS}} - \vec s$ and $\vec { m_{iP}} - \vec p$ for each telescope $i$. To do so, the targets are imaged in the $H$ band on the acquisition camera, and the sources are maintained on the fibers thanks to the tri-axis piezo mirror situated in the pupil plane of the instrument. The second terms to minimize are the mismatch between the imaging baseline limit-points and the metrology end-points: i.e., $\overrightarrow{\vec { I_{iS}M_i}}$ and $\overrightarrow{\vec { I_{iP} M_i}}$. To do so, four light sources (emitting at a wavelength of 1.2\,$\mu$m) are placed on the spider arms of each telescope. They are imaged through a 2x2 lenslet Shack-Hartmann on the acquisition camera. The low order wavefront sensor allows to derive both a lateral (tip-tilt) and longitudinal (focus)  error of the position of the pupil.  The lateral position is controlled thanks to a pupil actuator mirror situated in a focal plane inside the GRAVITY cryostat. The longitudinal error is corrected thanks to the VLTI delay-line variable curvature mirrors (VCM).

With this setup, we can position the imaging baseline limit-points close to the limit-points of the two other baselines. Fig.~\ref{figGRAV} gives a schematic representation of the expected position of the WAB, NAB, IMB$_1$, and IMB$_2$ limit-points in telescope coordinates.

Of all limit-points, the WAB limit points, the pivot points of the telescopes, are -- in theory -- the easiest to locate in geographical coordinates. They
are the center of rotation of the telescope, which can be  determined by external measurements. However, issues can come from bending of the structure 
which may hinder the notion of a fix pivot point.

 On the other hand, the problem of the NAB and IMB limit points are that they are defined in
telescope frame coordinates. Note that because of the telescope optical train (notably M2), the telescope frame
coordinates are not necessarily collinear to the celestial frame of the object: the telescope reference frame can be
several arcsec away from the direction of the target. The terrestrial coordinates are defined as ($u$,$v$,$w$). The telescope
coordinates are defined as ($x$,$y$,$z$), with (0,0,0) the coordinate of the pivot point, ($x$,$y$) the transverse
coordinates, and ($z$) the longitudinal coordinates.

On the UTs, the metrology measurements will be obtained thanks to 4 photodiodes positioned on the spider arms, symmetrically around the pupil. The $(x_d,y_d)$ coordinates are: (1.945,1.692), (-1.945,1.692), (1.945,-1.692) and (-1.945,-1.692) in meters. The positions were chosen to cancel any small defocus in the non-common path (Section~\ref{secGHO}). The metrology measurement $\Phi_M$ will be the mean of the four phase measurements. Hence, the resulting limit point of the NAB will be the
mean position of the four metrology sensors. It gives a NAB limit-point close to the center of the beam ($|<x_d,y_d>| \approx 0$). In longitudinal coordinates, the spider arms are at
$z\approx 10$\,m above the pivot point.

The uncertainty of the position of the imaging baseline limit-points is larger due to the relay optics between the cryostat and the telescope pupil. As a result, the effective pupils can move with respect to the telescope pupil. However, on GRAVITY, the internal pupil tracking system will stabilize the effective pupils close to the telescope pupil. The imaging limit points are expected to be at $|x,y| = 0 \pm 4$\,cm from the central optical axis. The tracking  in the longitudinal direction 
should also be able to perform with an expected accuracy of $z = 10$\,km between the metrology end-points and the imaging baseline limit-points \footnote{$z = 10$\,km in the 8 meter beam is equivalent to $z = 1$\,m in the 80\,mm beam in the VLTI laboratory.}.

\subsection{GRAVITY vulnerability to high-order aberrations}
\label{secGHO}

The two incoming stellar fields $E_p(x,y)$ and  $E_s(x,y)$  are defined over the eight meters of the telescope pupil except for the central 1.116 meter in diameter due to secondary mirror obscuration (0.1395 of the pupil radius). Their phases are affected by both the atmospheric turbulence and the optical aberrations in the optical train. 

The electromagnetic fields accepted by the fibers are also defined in the pupil: $E_{Fp}(x,y)$ and  $E_{Fs}(x,y)$. These fields have a near-Gaussian profile whose amplitude can be approximated by:
 \begin{equation}
 E_{Fp}(x,y) = E_{Fs}(x,y)= \exp\left(-\frac{x^2+y^2}{2\sigma_F^2}\right) 
 \end{equation}
where  $\sigma_F$ was chosen to maximize the injected flux at optimal correction (for a perfectly flat incoming wavefront). $\sigma_F =0.644573$ in unit of pupil radius. The phases of the $E_{Fs}$ and $E_{Fp}$ electric fields are only important relative to the phase of the incoming stellar fields ($E_s$ and $E_p$ respectively), so we suppose the phases of the acceptance fields equal to zero. The phase errors resulting from these fields are therefore:
\begin{eqnarray}
\delta \Phi_{\rm stellar}&=&\arg \left(  \iint \limits_{0.1395<\rho<1}  \!    E_{s1}(x,y)  \exp\left(-\frac{x^2+y^2}{2\sigma_F^2}\right)    \,  \mathrm dx \mathrm dy     \right) \nonumber \\
	&&-\arg \left(  \iint \limits_{0.1395<\rho<1}  \!    E_{s2}(x,y)    \exp\left(-\frac{x^2+y^2}{2\sigma_F^2}\right)  \,  \mathrm dx \mathrm dy     \right) \nonumber \\
&&-\arg \left(  \iint \limits_{0.1395<\rho<1}  \!    E_{p1}(x,y)    \exp\left(-\frac{x^2+y^2}{2\sigma_F^2}\right)  \,  \mathrm dx \mathrm dy     \right) \nonumber \\
&&+\arg \left(  \iint \limits_{0.1395<\rho<1}  \!    E_{p2}(x,y)    \exp\left(-\frac{x^2+y^2}{2\sigma_F^2}\right)  \,  \mathrm dx \mathrm dy     \right) \nonumber
\end{eqnarray}
where $\rho=\sqrt{x^2+y^2}$.

GRAVITY will have four metrology receivers on each telescope. According to Eq.~(\ref{eq:phimet}), the effect of internal optical aberrations is therefore:
\begin{equation}
\delta \Phi_{\rm met}=\frac{1}{4} \sum_{d=1}^4    \!  \sum_n \left[   (k^n_{s1} - k^n_{p1}-k^n_{s2} + k^n_{p2})\ Z^n(x_d,y_d)  \right]
\end{equation}
where $(x_d,y_d)$ are the positions of the four metrology sensors, and $k^n_{s1}$, $k^n_{p1}$, $k^n_{s2}$, $k^n_{p2}$ the amplitude of the aberrations affecting each one of the four metrology beams.

The four $(x_d,y_d)$ sensors coordinates  can be chosen so that optical aberrations affecting both stellar and metrology beams can be fully compensated : $\delta \Phi_{\rm stellar}\lambda_{s} + \delta \Phi_{\rm met}\lambda_{m} = 0$.
Because of the limited number of metrology receivers, the number of Zernike polynomials whose effect can be nullified is limited. To cancel a polynomial $n$, the $(x_d,y_d)$ coordinates must satisfy the relation:
\begin{equation}
\frac{1}{4} \sum_{d=1}^4     Z^n(x_d,y_d)  = \frac{
   \iint \limits_{0.1395<\rho<1}  \!  Z^n(x,y)  \exp\left(-\frac{x^2+y^2}{2\sigma_F^2}\right)     \,  \mathrm dx \mathrm dy    }{
   \iint \limits_{0.1395<\rho<1}  \!   \exp\left(-\frac{x^2+y^2}{2\sigma_F^2}\right)     \,  \mathrm dx \mathrm dy    } \ ,
\end{equation}
in the approximation of low amplitude phase errors.

The choice made for GRAVITY was to position the metrology sensors to ensure that piston, tip, tilt and defocus are canceled by the metrology.
For example, the fourth polynomial,
$Z_4(x,y)=2\sqrt 3(x^2+y^2)-\sqrt 3$ \citep[defocus,][]{1976JOSA...66..207N}, puts a constraint on the distance between the center of the pupil to the position of the sensors:
\begin{equation}
\rho_d^2  =\frac{
   \int_{\rho=0.1395}^1  \!   2\pi \rho^3\exp\left(-\frac{\rho^2}{2\sigma_F^2}\right)    \,  \mathrm d\rho }{  
   \int_{\rho=0.1395}^1  \!   2\pi \rho \exp\left(-\frac{\rho^2}{2\sigma_F^2}\right)   \,  \mathrm d\rho }
\end{equation}
gives  $\rho_d=0.644573$ in unit of pupil radius\footnote{$\rho_d=\sigma_F$, a consequence of injection optimisation}.
However, depending on the position of the metrology sensor, the influence of the other Zernike polynomials in the optical train can be amplified. The idea for GRAVITY was to cancel the aberrations which can be caused by a simple displacement of the single-mode fibers in the focal plane. The influence of the higher order polynomials is investigated numerically in Section~\ref{secNCPi}.

\subsection{Polarization in GRAVITY}

Much of the preceding discussion has revolved around the {\em spatial} properties of the GRAVITY instrument associated with the VLTI optical train, and their differential effects on science and reference stars, and the metrology light. The discussion would not be complete without addressing also the {\em polarization} states of light. The combined optical train of the VLTI and the GRAVITY instrument is birefringent. This causes phase errors in two ways. First, the differential phase between the metrology beams for the science and reference stars is impacted (in any telescope) if they do not enter the optical train with identical polarization orientation. For the expected magnitude of birefringence, this phase error can be made negligible with an easily achievable alignment accuracy. A second, more subtle effect, results from the combination of a)~{\em differential} birefringence between optical trains; b)~different polarization states of the science and reference starlights. In fact, the light from gas orbiting the galactic center black hole (a prime science target for GRAVITY) is known to be partially linearly polarized. 

A detailed account is beyond the scope of the present paper, and will be given in a forthcoming publication (Lazareff et al 2014, in preparation). We plan to operate with the assumption of an unpolarized reference star, and a partially {\em linearly} polarized science target. Under that condition, fringes detected in a linear polarization (at output) perpendicular to either eigenvector are immune from the phase error. Then, the operation can be summarized as follows: i) define (in our particular context) eigenvectors of the optical train as linear polarizations (at input) that result in linear polarization at output. ii) Rotate the polarization controllers of GRAVITY to realize the orthogonality condition. Because the birefringence of each mirror train is a function of the pointing parameters, this alignment condition will be determined {\em in situ} within a short time interval from the actual observation, or according to polarization modeling acquired during dedicated campaigns.

\section{Astrometric error budget}
\label{secerror}

\begin{table*}
\caption{GRAVITY astrometric error budget (between UTs, for a single baseline)}        
\label{table:1}     
\centering                      
\begin{tabular}{l c c c c}     
\hline\hline            
Error label & Cause & Amplitude considered & Consequence & $\mu$as\tablefootmark{a} \\  
\hline                     
&\bf $\vec { B_{\rm NAB}}$ Error  \\
$\overrightarrow{\vec { LM}}$ in terrestrial coordinates & Physical pointing of the telescope & $\pm 10\arcsec$ & Baseline error 0.5\,mm & 5 \\
$\vec { B_{\rm WAB}}$ stability  & Paranal structural stability & 0.5\,mm  & Baseline error 0.5\,mm & 5\\
Terrestrial to Sidereal reference frame & LST time during observation & 1/15\,s  & Baseline error 0.5\,mm & 5\\
\hline          
&\bf Pupil Error\\
Lateral Pupil  & $ (\vec  m - \vec  s) \cdot (\overrightarrow{\vec  {IM}} \perp z)$ &  10mas -- 4cm & OPD error of 2nm & 4 $\sqrt 4$ \\
Longitudinal Pupil  & $ (\vec m - \vec  s) \cdot (\overrightarrow{\vec  {IM}} \parallel z)$ &  10mas  -- 10km & OPD error of 10pm & 0.02$\sqrt 4$  \\
\hline
&\bf Dispersion Error \\
Dispersion of air & $\Delta_{air} \cdot(n^{\lambda_m}_{air}-n^{\lambda_s}_{air})$ & $1\,$mm -- $8\times 10 ^{-8}$ & OPD error of 80\,pm & 0.16 \\
Differential length of fiber &   $\Delta_{fiber} \cdot(n^{\lambda_m}_{fiber}-n^{\lambda_s}_{fiber})$ &  $10\,\mu$m -- $2\times 10 ^{-4}$ & OPD error of 2\,nm  & 4 \\
\hline
&\bf High order terms Error \\
Non Common path aberrations & details in Table~\ref{table:2} && OPD error of 3.7\,nm  &  7.4$\sqrt 2$  \\
Common path aberrations & details in Table~\ref{table:3} && OPD error of 0.9\,nm  &  1.8$\sqrt 2$ \\
\hline
\bf TOTAL & &&& {\bf  16.5\tablefootmark{b}} \\
\hline 
\end{tabular}
 \tablefoot{
\tablefoottext{a}{Accuracy obtained between two targets separated by 1 arcsec, projected on a 100\ m baseline}
\tablefoottext{b}{Using all six baselines, and needing astrometry on the two coordinate axis, the final accuracy would reach $16.5\sqrt{2/6}={ 9.5}\ \mu$as }
}
\end{table*}

The following sections derive GRAVITY's astrometric error budget from the expression of the astrometric measurement established from Eqs.~(\ref{eqOPDtotal}) and~(\ref{eqOPDerr}). The goal is to estimate the bias that may affect the measurement of $\vec  s - \vec p $. Section~\ref{secNABi}  considers the imprecisions on the $\vec { B_{\rm NAB}}$ baseline vector. Section~\ref{secIMBi}  focuses on the combined effect of tip-tilt error (the $\vec { m_s}-\vec  s$ and $\vec { m_p}-\vec  p$ terms) with a pupil matching error ($\overrightarrow{\vec { IM}} \neq \overrightarrow{\vec 0}$). Section~\ref{secDispi}  quantifies the errors brought by the dispersion in the interferometer (The $D(\lambda_{m},\lambda_s)$ value). Last, in Section~\ref{secPhai}, the errors caused by the high order aberrations are computed: $\delta \Phi_{\rm stellar}$ and $\delta \Phi_{\rm met}$.

\subsection{NAB determination in celestial coordinates}
\label{secNABi}

The fact that the limit points of the metrology are not at the same position as the pivot points ($\overrightarrow{\vec { LM}} \ne \overrightarrow{\vec  0}$ in
Eq.~(\ref{OPDIMG})) mostly raises a problem of reference systems. The
WAB
baseline is defined in terrestrial coordinates, and fixed in this frame of reference. On the other hand, the narrow angle baseline
 is defined in a different frame system: the telescope coordinates. Finally, the stellar
object, $ (\vec s - \vec p)$ is set in celestial coordinates.

The effect of the NAB on the OPD measurement appears in the first component of Eq.~(\ref{OPDIMGL}) : $(\vec  s - \vec  p) . (\vec { B_{\rm WAB}} + \overrightarrow{\vec {M_1L_1}} - \overrightarrow{\vec {M_2L_2}})$.
Uncertainties come from:
\begin{itemize}
 \item How well the $\overrightarrow{\vec { LM}}$  vectors are known in terrestrial coordinates;
 \item How well the $\vec { B_{\rm WAB}}$ is known in terrestrial coordinates;
 \item How well the terrestrial coordinates are known with respect to sideral coordinates.
\end{itemize}

The first  item is an issue of telescope pointing, telescope bending, telescope dilatation. The
relative orientation between telescope and terrestrial coordinates can be obtained to a high precision (and high repeatability) thanks to telescope
encoders.  Note that the physical and optical pointing of the telescope are not necessarily the same. What matters to determine the $\overrightarrow{\vec { LM}}$ vector in terrestrial coordinates is
 the physical pointing of the telescope. It is expected to have a reproducibility of roughly 10 arcsec. Hence, for GRAVITY, the
error caused by the projection of the 10\,m distance between the two limit points is of the order of $10\,\rm m \times 10 \arcsec =
0.5$\,mm.

The second item is an issue of stability and measurement. Over time, if stable enough, the WAB can be averaged to a very high precision. The stability is a structural issue, inherent to the infrastructure of the interferometer. It is expected to be stable at the sub-millimetre level (except perhaps in the event of a big earthquake). In the error budget, we have assumed an uncertainty of 0.5\,mm in the WAB.

The relative orientation between the reference frame of the stellar target and the terrestrial coordinates depends on i) the knowledge of the target coordinates, and ii) the local stellar time (LST). We assume that time accuracy will be the limiting factor. Not because the LST is not precisely known, but simply because integration times on the science camera are of the order of several seconds. Putting a precise time-stamp on the data will therefore be difficult, especially since it would have to be weighted by injection coefficients during the exposure. We estimate our uncertainties to 1/15\ s, meaning $1\arcsec$ on sky, knowing that the earth rotate by 15 arcsec per second. The  error caused by an arcsec on the projection of a 100\,m baseline is of the order of $0.5\,$mm (decreasing with the proximity to the zenith).

With respect to a 100\,m baseline, 0.5\,mm error in baseline will correspond to an error in the astrometric accuracy
of 5ppm. Hence $5\ \mu$as for two stars separated by $1\arcsec$. The NAB indetermination factors are summarized in Table~\ref{table:1}.

\subsection{IMB to NAB mismatch}
\label{secIMBi}

The IMB has an effect on the astrometric measurement only if i) the metrology beam is not exactly pointing toward the stellar source ($\vec 
m-\vec  s \neq
\vec  0$) {\em and} ii) the limit points of the IMB and the NAB do not agree ($\overrightarrow{\vec  {IM}}\neq \overrightarrow{\vec 0}$). The effect on the OPD measurements is an additional term for each telescope: $(\vec { m_{iS}} - \vec s) \cdot \overrightarrow{\vec {I_{iS}M_S}} - (\vec { m_{iP}} - \vec p) \cdot \overrightarrow{\vec {I_{iP}M_P}}$, with $i$ the telescope number.
Although it is technically possible to model its influence, due to the complexity of this term, it is better to try to nullify it.  This is achieved in GRAVITY thanks to the pupil tracking camera mentioned in Section~\ref{BaseSpec}. The residuals are taken into account in the error budget.

The pupil tracking camera will track the pupil both on the longitudinal and lateral directions. But due to its design, the longitudinal accuracy is consequently larger than lateral. With the pupil tracking activated, the pupil vector $\overrightarrow{\vec {IM}}$ is expected to be within 10\,km in the longitudinal and 4\,cm in the lateral direction ($z= 0 \pm 10\,$km and $|x,y| = 0 \pm 4\,$cm).

In the GRAVITY design (as shown in Fig.~\ref{figGRAV}) the metrology beam is injected inside the IO beam combiner. As a result, by optimizing the injection of the stellar flux inside the fibers, GRAVITY will also make sure that the metrology beam has a propagation direction near-parallel to the stellar light (but with opposite direction). Hence for optimum injection,  $\vec m-\vec s = \vec 0$.
Practically, GRAVITY acquisition camera should stabilize the injection during observation enough  to be able to reach $|\vec  m-\vec  s | \leq 0.2 \lambda/D$, or $\approx 10$\,mas at $\lambda=2.2 \mu\,$m.

We can split the OPD error induced by the IMB to NAB mismatch into two terms: longitudinal and lateral. The lateral contribution is the 4\,cm of the pupil displacement projected toward the normalized $(\vec m-\vec s)$ vector: $\sin ( 10\,{\rm mas}) \times 4$\,cm$=2$\,nm. The longitudinal
contribution will be $[1-\cos ( 10\,{\rm mas}) ] \times 10$\,km$=0.01$\,nm. Related to the total OPD expected between
two target separated by 1'' observed with a baseline of 100\,m, it means that the IMB to NAB mismatching will result,
for GRAVITY, in an additional astrometric error of $4\ \mu$as. To account for each 4 terms, this value is multiplied by $\sqrt 4$ in  Table~\ref{table:1}.

\subsection{Dispersion}
\label{secDispi}

According to Eq.~(\ref{eqDisp}) the dispersion term wrote:
\begin{equation}
 D(\lambda_{m},\lambda_s)= \Delta_{air} \cdot(n^{\lambda_m}_{air}-n^{\lambda_s}_{air}) + \Delta_{fiber} \cdot (n^{\lambda_m}_{fiber}-n^{\lambda_s}_{fiber})\,,
\end{equation}
where $\Delta_{air}$ and $\Delta_{fiber}$ are the differential delays in air and glass respectively. Ideally, the differential path is entirely compensated in fibers: $\Delta_{air} \approx 0$ and $\Delta_{fiber} \approx \delta OPD_M$. This is not exactly the case due to non-common path in air or vacuum. 
The main errors come from uncertainties in the physical length of fiber ($\Delta_{fiber}$) and of air ($\Delta_{air}$). But uncertainties in the knowledge of the refractive indices ($n^{\lambda_m}-n^{\lambda_s}$) can also be a problem, since they depend on environmental conditions.

Following \citet{1967ApOpt...6...51O} formula, we computed the dispersion of air for standard atmospheric conditions. It gives:
\begin{equation}
n^{1.9 \mu m}_{air}-n^{2.2 \mu m}_{air}= 8\times 10^{-8}\,,
\end{equation} 
hence a dispersion effect of 80pm for $\Delta_{air} =1mm$. This is a conservative upper limit, since the two stellar beams share the same optics up to the cryostat (the star separator is inside the cryostat). In the data reduction software, air dispersion will be therefore neglected.

The fluoride fibers are however a lot more dispersive \citep{1995A&A...293..278C}. Thus, their effect cannot be neglected. Thanks to the fact that the fibers are inside the cryostat, we assume that the dispersion in the fibers will not change, and that it can be precisely modeled. First laboratory results show dispersion coefficients of the order of 
\begin{equation}n^{1.9 \mu m}_{fiber}-n^{2.2 \mu m}_{fiber}= 2\times 10^{-4}\,.
\end{equation} 
 This is in accordance with  \citet{1995A&A...293..278C}. For a physical length $\Delta_{fiber} =1mm$, the correction term to the OPD is therefore of the order of 200\,nm. With a knowledge of  the fiber stretching of the differential delay lines of $\sigma(\Delta_{fiber}) = 10\,\mu$m, we can model the $D(\lambda_{m},\lambda_s)$ to a precision of 2\ nm.

\subsection{Optical aberrations}
\label{secPhai}

These aberrations can be split into two distinct categories:
\begin{itemize}
\item The common-path aberrations which affect both stellar signals. This is the case of atmospheric perturbations, partially corrected by the AO on the UTs. 
\item The non-common-path aberrations which affect only one of the stars. This is typically a consequence of a beam-walk effect, when the two stellar beams do not have the same imprint on the optical surfaces. 
\end{itemize}
We assume that the metrology will monitor the non-common-path optical aberrations, but not the common-path aberrations. We neglect isoplanetism, as being out of scope of this paper\footnote{For PRIMA, analytical models predict a precision of 10\ $\mu$as for one hour of observations between two targets separated by 10 arcsec \citep{2013A&A...551A..52S}.}.

\subsubsection{Non-common-path aberrations -- beam-walk defects}
\label{secNCPi}

\begin{table}
\caption{Error caused by non-common-path aberrations}        
\label{table:2}     
\centering                      
\begin{tabular}{l c r r c}     
\hline\hline            \vspace{-.25cm}\\
Zernike\#\tablefootmark{a} & rms\tablefootmark{b}  & $\delta \Phi_{\rm stellar}\frac{\lambda_s}{2\pi}$ & $\delta \Phi_{\rm met}\frac{\lambda_m}{2\pi}$ & $\delta\Phi_{\rm stellar}\frac{\lambda_s}{2\pi} +\delta \Phi_{\rm met}\frac{\lambda_m}{2\pi}$\\
\hline
4\tablefootmark{c}& 8\,nm &-2.3\,nm & -2.3\,nm & 0.0\,nm \\
6\tablefootmark{d}& 8\,nm &0.0\,nm & 1.1\,nm & -1.1\,nm \\
11\tablefootmark{e}& 2\,nm &-0.1\,nm & -2.7\,nm & 2.6\,nm \\
12 & 2\,nm &0.0\,nm & -0.6\,nm & 0.6\,nm \\
14 & 2\,nm &0.0\,nm & -1.4\,nm & 1.4\,nm \\
22\tablefootmark{f}& 1\,nm &0.1\,nm & 0.7\,nm & -0.7\,nm \\
24 & 1\,nm &0.0\,nm & 0.1\,nm & -0.1\,nm \\
26 & 1\,nm &0.0\,nm & 1.6\,nm & -1.6\,nm \\
28 & 1\,nm &0.0\,nm & -0.1\,nm & 0.1\,nm \\
\hline 
Total & & & & 3.7\,nm \\
\hline
\end{tabular}
 \tablefoot{
\tablefoottext{a}{Zernike numbering as defined by \citet{1976JOSA...66..207N}} 
\tablefoottext{b}{Amplitude of the aberration due to the Zernike in nm}
\tablefoottext{c}{Defocus}
\tablefoottext{d}{Astigmatism}
\tablefoottext{e}{3rd order spherical}
\tablefoottext{f}{5th order spherical}
}
\end{table}

The goal of the metrology is to monitor and account for the non-common optical path length between the two beam combiners. By construction, the metrology laser does perfectly probe any piston induced in the optical train. Higher order are not necessarily measured as well. To simulate the effect of non-common-path aberrations, we consider the influence of Zernike polynomial ($n$) individually, affecting only one beam of the metrology ($k^n_{s2}=k^n_{p1}=k^n_{p2}=0$). Then:
\begin{equation}
\delta \Phi_{\rm met}=\frac{1}{4} \sum_{d=1}^4    \!   k^n_{s1} \ Z^n(x_d,y_d) \ .
\end{equation}
The same aberration also affects the stellar light, but with a negative sign:
\begin{equation}
E_{s1}(x,y)=\exp (-\imath\, k^n_{s1} Z^n(x,y)) 
\end{equation}
hence:
\begin{equation}
\delta \Phi_{\rm stellar}=\arg \left(  \iint \limits_{\rm pupil}  \!  \exp (-\imath\, k^n_{s1} Z^n(x,y))    \exp\left(-\frac{x^2+y^2}{2\sigma_F^2}\right)    \,  \mathrm dx \mathrm dy     \right)
\end{equation}
giving an OPD error according to Eq~(\ref{eqOPDerr}) of $(\delta\Phi_{\rm stellar}\frac{\lambda_s}{2\pi} +\delta \Phi_{\rm met}\frac{\lambda_m}{2\pi})$ in nanometers.

The amplitude of the non common path aberrations were measured during two test campaigns at Paranal in August and October 2010. They were used to validate the metrology principle. Two coherent beams were overlapped at an angle corresponding to the angular separation of
 two stars in the field of view of  GRAVITY. The two planar waves interfere
to create a characteristic fringe pattern inside the VLTI lab. This fringe pattern
was projected into the VLTI, including beam expanders and delay lines, onto the pupil of UT4. Aberrations were measured  on the M2 mirror from the scattered light, observed with a commercial IR camera mounted to the telescope structure.
 We found that deviations occurring at the edge of the pupil are
smaller than 8nm / 2nm for a time base of 30 minutes (for quadratic and forth order respectively, negligible above).

In Table~\ref{table:2} are summarized the errors on the OPD measurement for each Zernike polynomials up to the seventh order (Zernike $\#\leq 36$). Due to the point-symmetry of the fundamental mode of the fibers, as well as of the metrology sensor, most of the polynomials do not bias the measurement. Only the ones affecting the phase are shown in Table~\ref{table:2}. The variance considered for each one of these Zernike is 8\ nm for the quadratic order, 2\ nm for the forth order, and 1\ nm for the sixth order. The quadratic sum of all these terms is shown in the last row, which must be multiplied by $\sqrt 2$ to account for errors induced by the two arms of the interferometer.

\subsubsection{Common-path aberrations -- atmospheric perturbations}
\label{secCPi}

\begin{table}
\caption{Error caused by atmospheric perturbations}        
\label{table:3}     
\centering                      
\begin{tabular}{c r r r}     
\hline\hline            
$\vec { \Delta p}$ & UTs  & UTs\tablefootmark{a} & ATs \tablefootmark{a} \\
\hline
 $0.0 \lambda/D$&    0 nm &    0 nm &    0 nm \\
 $0.1 \lambda/D$&    4 nm &   96 nm &   53 nm \\
 $0.2 \lambda/D$&    9 nm &  173 nm &  123 nm \\
 $0.3 \lambda/D$&   15 nm &  240 nm &  190 nm \\
 $0.4 \lambda/D$&   22 nm &  $> 300$ nm\tablefootmark{b} &  243 nm \\
 $0.5 \lambda/D$ &   32 nm &  $> 300$ nm\tablefootmark{b} &  $> 300$ nm\tablefootmark{b} \\
\end{tabular}
 \tablefoot{
\tablefoottext{a}{without AO} 
\tablefoottext{b}{lower limit due to $2\pi$ wrapping} 
}
\end{table}

The atmosphere perturbations affect the phase of the electromagnetic field coming from the stars. If we neglect the influence of anisoplanetism, we can consider a similar electric field coming from both stars $E_p(x,y)\propto E_s(x,y) \propto E(x,y)$. The atmospheric perturbations are therefore common-path aberrations. They are not probed by the metrology ($\delta \Phi_{\rm met}=0$) and affect the optical path difference as:
\begin{eqnarray}
\delta \Phi_{\rm stellar} &= & \arg \left(  \iint \limits_{pupil}  \!    E(x,y) E^*_{Fp}(x,y) \,  \mathrm dx \mathrm dy     \right)
\nonumber \\ && - \arg \left(  \iint \limits_{pupil}  \!    E(x,y) E^*_{Fs}(x,y) \,  \mathrm dx \mathrm dy     \right)
\end{eqnarray}
where $E^*_{Fp}(x,y)$ and $ E^*_{Fs}(x,y)$ are the accepted mode of the fibers. The effect of atmospheric turbulence is here computed for one telescope only, since the atmosphere on the two telescopes can be considered as uncorrelated.

In the absence of any non common path aberrations  ($ E^*_{Fp}(x,y) = E^*_{Fs}(x,y) $), common path aberrations do not change the measured optical path length: $\delta \Phi_{\rm stellar} =0$. It is therefore necessary to combine them with some level of field and/or pupil offset. Considering a field offset, ie, a tip-tilt error, we used the following equations to simulate the acceptance fields:
\begin{eqnarray}
E_{Fp}(x,y) &=&\exp\left(2\pi \imath \frac{\vec { \Delta p} \cdot x}{\lambda_s}-\frac{x^2+y^2}{2\sigma_F^2} \right)\\
E_{Fs}(x,y) &=&\exp\left(-\frac{x^2+y^2}{2\sigma_F^2} \right)
\end{eqnarray}
where $\vec { \Delta p}$ is the tip-tilt error. Such tip-tilt error will necessarily be present because the alignment of the fiber cannot be perfectly on the star. For $E(x,y)$, we used the YAO adaptive optics simulator\footnote{http://frigaut.github.io/yao/index.html} to generate arrays of phase masks for three different conditions: i) observation with the UTs, with the AO locked on a Kmag=7 off-axis reference star, ii) observation with the UTs without AO, and iii) observation with the ATs\footnote{As yet, the ATs are not equipped with an AO system}. The seeing conditions are average ($r_0=25$\ cm at $\lambda = 0.5 \mu$m).
The results are presented in Table~\ref{table:3} for tip-tilt errors from  0 to $0.5\lambda/D$.

The OPD errors average with time. However, the coherence time of the common path aberrations depends highly on the seeing conditions, wind speed, etc... A rough, conservative, solution is to consider a coherence time of 100\ ms, hence, a reduction by a factor 10 for a 10 second observation. In the case of the $0.2\lambda/D$ stated in Section~\ref{secIMBi}, it means a 0.9\ nm OPD error on the UTs, and 12.3\ nm error on the ATs. A factor $\sqrt 2$ is necessary to account for the atmospheric perturbations on each telescope. This conservative value is reported in Table~\ref{table:1}.

It is important to note that the fiber positioning (responsible for the tip-tilt error) is crucial to an operation without AO. On the ATs, if the alignment is below $0.5\lambda/D$, the phase measurement will wrap over $2\pi$, preventing averaging of the error, and thus any realiable astrometric 

\section{Discussion \& Conclusion}
\label{secconclusion}

\subsection{15 years of experimentation}

The predominantly theoretical exercise of this paper is supported by 15 years of experimentation carried out on various interferometers. The first astrometric observations with a long baseline interferometer were achieved around 1999 on the Palomar Testbed Interferometer \citep{1999ApJ...510..505C}, based on the theoretical work of \citet{1992A&A...262..353S}. Using small siderostats on 100 meter-long baselines, the instrument reached 160\ $\mu$as precision measurements on bright pairs \citep{1999AAS...195.8714S}. PTI was a technology demonstrator in preparation to the four 1.8 m telescopes planned to be added to the Keck Interferometer. The Keck Outrigger astrometry project was the first to tackle the issue of implementing astrometric baselines on large telescopes. The concept foresees to transfer the metrology end-point to primary space by a dedicated baseline monitor \citep{2004SPIE.5491.1649H}. However, the project was cancelled in 2006 before seeing first light. Started in 2008, and even though it never managed an on-sky astrometric demonstration, the ASTRA project of astrometric extension of the two 10 meter Keck Interferometer was the occasion to further study the implementation of metrologies for astrometric interferometer on large telescopes.

The design of PRIMA \citep{2008NewAR..52..199D}, the VLTI equivalent to ASTRA and the Keck Outriggers, started in 2000. In 2011, during its first on-sky observations, PRIMA achieved an astrometric precision of 30\ $\mu$as, but the astrometric accuracy was limited to about 3 mas, mainly because the metrology endpoints were located at the telescopes Coude foci, far from primary space \citep{2013A&A...551A..52S}.
A second on-sky demonstration run in 2013, with the metrology extended to the telescope secondary mirror, resulted in an astrometric accuracy of order 100\ $\mu$as.

The GRAVITY metrology \citep{2012SPIE.8445E..1OG} follows a slightly different concept than the previous experiments. Instead of measuring in double pass the optical path difference between two telescopes for each star, as done for PRIMA and ASTRA, the GRAVITY metrology measures the optical path difference between the two beams for each telescope in single pass. And other than in PRIMA and ASTRA, the GRAVITY metrology end points are mounted above the primary mirror on the telescope spider, therefore covering the full optical path and providing a physical realization of the narrow angle baseline endpoints. A second laser is send back from the telescope to actively track the pupil on the metrology endpoints \citep{2012SPIE.8445E..34A}.

\subsection{On the GRAVITY astrometric error budget}

A sophisticated error budget was a central design driver for GRAVITY. The idea is to cancel as many error terms as possible thanks to technical choices. For example,  the metrology is sent up to the spider arms, above M1. It is located in geographical coordinates to minimize error when transposing the physical narrow angle baseline to the sidereal reference frame (Section~\ref{secNABi}). The metrology is also send by the same single-mode fibers that accept the star light. The difference in  propagation direction between the metrology and the starlight is therefore minimized (Section~\ref{secIMBi}).

The necessity of \textbf{an active control over the pupils} (the imaging baseline limit-points) is one of the main requirement given by the error budget. The imaging baselines introduce an error term if the instrument pupil and focal plane are not correctly stabilized to a given position (Section~\ref{secIMB}). This issue cannot be canceled out by technical design, so GRAVITY includes inside its own cryostat a pupil and field tracking system. This system  controls in real time the position of the instrument pupil with respect to four laser beacons on the spider arms of the telescopes. It also monitors the position of the stellar objects to keep them accurately positioned (within $0.2\lambda/D$) on the fibers.

\textbf{Field tracking is also a fundamental issue} to minimize the influence of common-path aberrations. At first sight, common path aberrations (eg., atmospheric perturbations) do not matter. This is not the case if there is an additional error on the field tracking: it is the cross-term between field error and atmospheric perturbation which matters. In
section~\ref{secCPi}, we showed that atmospheric perturbations have the potential to be the major limit of the astrometric precision, depending on the level of star tracking. Moreover, we show that GRAVITY's accuracy will depend on the level of AO correction. Notably, the absence of AO on the ATs does put a stringent requirement on the affordable tip-tilt errors for the system to work (see Table~\ref{table:3}).

The main result of the error budget is that GRAVITY should fulfill its objective of $10\ \mu$as accuracy on the UTs if: i) the telescope pupils are positioned within the instrument to an accuracy of 0.5\% of their diameter (4\ cm on M1) and ii) the two stellar objects are tracked within $0.2\,\lambda/D$.

\begin{acknowledgements}
S.L. would like to thank J.D. Monnier for fruitful discussions about the impact of common-path aberrations.
\end{acknowledgements}


\end{document}